\documentclass[aps,prl,twocolumn,showpacs,superscriptaddress,floatfix]{revtex4-1}
\usepackage{amsmath,amssymb}
\usepackage{graphicx}
\usepackage{epsfig}
\usepackage{color}
\usepackage{rotating}
\usepackage{bm}
\usepackage{setspace}

\begin{document}
\title{Enhancement of superconductivity in electron-hole coexisting Sr$_{1-x}$Eu$_{x}$CuO$_{2+y}$ films}
\author{Hang Yan}
\author{Ze-Xian Deng}
\author{Xue-Qing Yu}
\author{Yan-Ling Xiong}
\author{Qun Zhu}
\affiliation{State Key Laboratory of Low-Dimensional Quantum Physics, Department of Physics, Tsinghua University, Beijing 100084, China}
\author{Ding Zhang}
\author{Can-Li Song}
\email[]{clsong07@mail.tsinghua.edu.cn}
\affiliation{State Key Laboratory of Low-Dimensional Quantum Physics, Department of Physics, Tsinghua University, Beijing 100084, China}
\affiliation{Frontier Science Center for Quantum Information, Beijing 100084, China}
\author{Xu-Cun Ma}
\email[]{xucunma@mail.tsinghua.edu.cn}
\affiliation{State Key Laboratory of Low-Dimensional Quantum Physics, Department of Physics, Tsinghua University, Beijing 100084, China}
\affiliation{Frontier Science Center for Quantum Information, Beijing 100084, China}
\author{Qi-Kun Xue}
\affiliation{State Key Laboratory of Low-Dimensional Quantum Physics, Department of Physics, Tsinghua University, Beijing 100084, China}
\affiliation{Frontier Science Center for Quantum Information, Beijing 100084, China}
\affiliation{Beijing Academy of Quantum Information Sciences, Beijing 100193, China}
\affiliation{Southern University of Science and Technology, Shenzhen 518055, China}

\begin{abstract}
We report transport measurements of infinite-layer cuprate Sr$_{1-x}$Eu$_{x}$CuO$_{2+y}$ films with controlled electron (by trivalent europium) and hole (by interstitial apical oxygen) carriers grown on SrTiO$_3$(001) with molecular beam epitaxy. An unusual enhancement of superconductivity upon moderate electron-hole coexistence in the films is found, which spans over the whole superconducting phase diagram and becomes more prominent in the underdoped regime. The superconductivity exhibits a two-dimensional nature with a thickness of approximately 5.2 nm, irrespective of the varying carriers, confirmed by angle-resolved magnetoresistance measurements and the Berzzinsky-Kosterlitz-Thouless transition. Nevertheless, the electron-hole coexistence enlarges the thermal activation energy of vortex motion that deviates obviously from the usual logarithmic evolution with the magnetic field. Our results offer a promising perspective to understand and enhance the high-temperature superconductivity in cuprates.
\end{abstract}

\maketitle
\begin {spacing}{1.05}
The coexistence of electrons and holes in solids brings about a plethora of extraordinary phenomena such as the exciton condensation, non-saturating magnetoresistance and unusual superconductivity \cite{Monney2011Exciton,kogar2017signatures,ali2014large,kamihara2008iron,si2016high,jindal2023coupled}. In particular, interactions between coexisting electron and hole bands at the Fermi level ($E_\textrm{F}$) have been widely proposed to be responsible for the high-temperature ($T_\textrm{c}$) superconductivity in iron pnictides \cite{Unconventional2008Mazin,Unconventional2008Kuroki,hirschfeld2011gap,Kontani2010Orbital}. For cuprate superconductors, however, the superconductivity has been commonly modeled with a single band of electrons or holes \cite{Zhang1988Effective, Lee2006doping, Scalapino2012common}. Yet, a few pieces of experimental evidence emerge and hint at a unique coexistence of electrons and holes across almost all cuprate superconductors \cite{Forro1990Hall,Ando2004Evolution,leboeuf2007electron,laliberte2011fermi,Dagan2004evidence,li2019hole,barivsic2022high}. This holds especially true for the electron-doped cuprates, because it turned out to be very challenging to remove all interstitial apical oxygens as unintentional hole dopants \cite{Armitage2010progress,Higgins2006Role, zhong2020direct,Molecular2020fan,Wang2020electronic,Yu2022percolative,Yan2023epitaxial}. Thus far, the major role played by the electron-hole coexistence (EHC) in the high-$T_\textrm{c}$ superconductivity of cuprates remains to be elucidated.

Infinite-layer cuprates stand out to be a promising compound to investigate this significant issue, because, in addition to the simplest structure, they can be both electron- and hole-doped without modifying the crystal motif \cite{zhong2020direct}. Moreover, the infinite-layer cuprates are isostructural to the recently discovered 112-type nicklate superconductors at ambient pressure, in which the EHC has been intensely discussed as well \cite{li2019superconductivity,Li2020superconducting,Zeng2020phase,Osada2020Phase, zeng2022superconductivity,Osada2020Phase,zeng2022superconductivity}. Thanks to a considerable solubility of the trivalent europium ions (Eu$^{3+}$) in SrCuO$_2$, infinite-layer Sr$_{1-x}$Eu$_{x}$CuO$_{2+y}$ (SECO) films over a wide electron doping range (with $x$ up to 0.25) have been recently prepared with a state-of-the-art molecular beam epitaxy (MBE) technique \cite{Yu2022percolative,Yan2023epitaxial}. Here, we report a sizable enhancement of $T_\textrm{c}$ upon co-doping a larger amount of interstitial apical oxygens (i.e.\ $y$) as hole dopants in the epitaxial SECO films [Fig.\ 1(a)]. The $T_\textrm{c}$ enhancements exist over the whole phase diagram and reach a maximum value of $\sim$ 29 K in the underdoped regime. Our Hall effect measurements suggest that the enhanced $T_\textrm{c}$ links intimately to a moderate matching of the coexisting electron and hole charge carriers. In addition, the angular dependence of the critical magnetic field and the observed Berezinskii-Kosterlitz-Thouless transition reveal a two-dimensional (2D) nature of the superconductivity.

High-quality SECO films were prepared on semi-insulating SrTiO$_3$(001) substrates with a home-built MBE apparatus equipped with an ozone delivery system for sample growth and a reflection high-energy electron diffraction (RHEED) for \textit{in situ} real-time analysis, as described in detail elsewhere \cite{Yu2022percolative,Yan2023epitaxial}. All samples studied have a fixed thickness of about 16.5 nm and were post-annealed in ultrahigh vacuum for one hour to achieve the optimal superconductivity \cite{Yan2023epitaxial}. These SECO samples were then transferred out of the MBE chamber for X-ray diffraction (XRD) analysis using the monochromatic Cu $K_{\alpha1}$ radiation with a wavelength of 1.5406 $\textrm{\AA}$. We measured the electrical resistivity, magnetoresistance and Hall coefficients ($R_\textrm{H}$) in a Quantum Design physical property measurement system (PPMS).
\end {spacing}

As evidenced before \cite{Yan2023epitaxial}, traces of long-$c$ phased SECO emerge as the (Sr + Eu)/Cu flux ratio exceeds a certain threshold of $\sim$ 1.1 during the MBE codeposition. Such a secondary phase is caused by oxygen over-stoichiometry at the apical sites of Cu and their ordering into the 2 $\times$ 2 superstructure \cite{zhong2020direct, leca2006superconducting}, as illustrated in Figs.\ 1(b) and 1(c). The apical oxygens serve as hole dopants and coexist with the intentional Eu dopants as electron doping [Fig.\ 1(a)]. We therefore label them as EHC SECO films and the electron-dominated ones involving no long-$c$ phase as ED SECO films. Figure 1(d) plots the temperature dependence of in-plane electrical resistivity $\rho_{ab}$ for two EHC films at $x \sim$ 0.133 and 0.141. Compared to the two ED samples with the same Eu doping level $x$, the $T_\textrm{c}$ is apparently enhanced for the EHC samples, albeit a twofold increase of the normal-state $\rho_{ab}$ just above $T_\textrm{c}$. For the $x \sim$ 0.133 sample, the $T_\textrm{c}$ onset increases more than threefold from 8.4 K to 29.0 K.

\begin{figure}[t]
\includegraphics[width=\columnwidth]{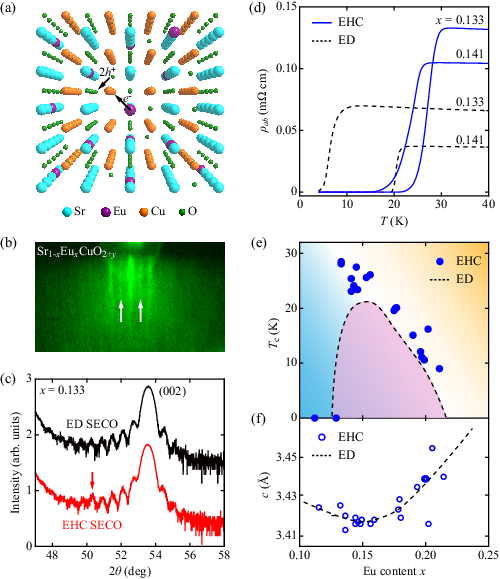}
\caption{(a) Schematic diagram of co-doped infinite-layer cuprates by partial substitutions of the trivalent cations Eu$^{3+}$ for Sr$^{2+}$ and the interstitial apical oxygens. (b) RHEED pattern in the EHC SECO sample ($x \sim$ 0.133) showing a (2 $\times$ 2) superstructure, highlighted by the two arrows, due to the presence of long-$c$ phase SECO. (c) XRD spectra around the (002) diffraction peak of the ED and EHC SECO films at the same Eu doping $x \sim$ 0.133. The red arrow marks the signal from the long-$c$ phase SECO with an out-of-plane lattice constant $c\sim$ 3.622 \AA. (d) In-plane electrical resistivity $\rho_{ab}$ for both ED and EHC SECO samples at different electron doping $x$, where $T_\textrm{c}$ is defined by the criterion of 90$\%$ normal-state resistivity before transition. (e, f) Phase diagram showing the variations of $T_\textrm{c}$ and $c$ for both ED (black dashes) and EHC (blue circles) SECO samples, respectively.
}
\end{figure}

\begin{figure}[h]
\includegraphics[width=\columnwidth]{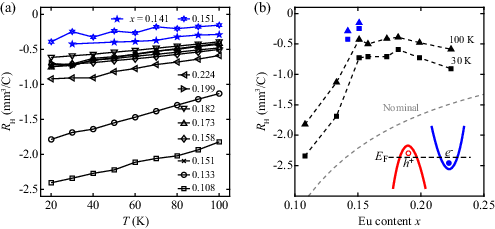}
\caption{(a) Hall coefficient $R_\textrm{H}$ for ED samples with $x$ ranging from 0.108 to 0.224 (black symbols) and two EHC samples with $x \sim$ 0.141 and 0.151 (blue symbols), plotted as a function of the temperature $T$. (b) Evolution of $R_\textrm{H}$ with the Eu content $x$ at $T$ = 30 K (squares) and $T$ = 100 K (triangles) for the ED (black symbols) and EHC (blue symbols) SECO films in (a). The grey dashed line corresponds to the theoretical $R_\textrm{H}$ for the nominal Eu content $x$ with a single-band model in which every Eu$^{3+}$ contributes one electron. Inserted in the lower right corner is a schematic two-band model where the $E_\textrm{F}$ intersects both electron-like (blue) and hole-like (red) pockets.
}
\end{figure}

We highlight that the $T_\textrm{c}$ enhancement occurs over the whole superconducting phase diagram universally rather than accidentally in the EHC samples, as compellingly revealed in Fig.\ 1(e). Here the dashed line reproduces the attainable $T_\textrm{c}$ maxima for the ED SECO samples at various Eu contents $x$ \cite{Yu2022percolative}, while the blue solid circles draw those of the EHC SECO samples. It should be emphasized that the width of superconducting transition nevertheless broadens to some extents in the EHC samples as compared to ED samples, which can be potentially attributed to increased scattering caused by apical oxygens and the presence of insulating long-$c$ phase SECO. This broadening largely reduces $T_\textrm{c0}$ in some SECO samples, to the extent that it can be lower than that observed in the ED samples. Since electrical transport measurements predominantly reflect the macroscopic bulk properties, and the $T_\textrm{c}$ onset is governed by the best superconducting path within the film, the enhanced $T_\textrm{c}$ onset in EHC samples is a direct indicator of improved superconductivity. Given that the EHC samples contain tiny traces of the long-$c$ phase SECO, the $T_\textrm{c}$ enhancement cannot be originated from the long-$c$ phase SECO or its interfacing with the ED SECO. In fact, the long-$c$ phase SECO is not superconducting at all, as revealed in Supplementary Fig.\ S1 \cite{supplementary}. At the same time, our findings are not reconcilable with a simple doping compensation between oppositely charged carriers by the trivalent Eu and apical oxygen. If so, the superconductivity should have been suppressed, other than most prominently enhanced in the underdoped regime [Fig.\ 1(e)]. Furthermore, we have calculated the out-of-plane lattice constants $c$ from the XRD measurements that present no systematic deviation from those of the ED samples [Fig.\ 1(f)]. This rules out possible structural distortions as the cause of the $T_\textrm{c}$ enhancement.

To provide more insight into the $T_\textrm{c}$ enhancement in the EHC samples, we show evolutions of the Hall coefficient $R_\textrm{H}$ with the temperature and Eu content $x$ for various SECO samples in Fig.\ 2. In all samples, there exists no apparent deviation of the Hall resistance $R_{xy}$ from linear dependence on magnetic field $\mu_0$$H$ up to 9 T at various temperatures [Supplementary Fig.\ S2] \cite{supplementary}. However, the extracted $R_\textrm{H}$ is unphysically smaller in magnitude than the theoretical value with single-band doping model, marked by a grey dashed line ($R_\textrm{H}$ $\varpropto -1/x$) in Fig.\ 2(b). This discrepancy becomes more pronounced close to the optimal Eu doping and in the EHC samples. We notice that the $R_\textrm{H}$ is always negative and shows no sign reversal at all temperatures and Eu contents. This excludes a possible Fermi surface reconstruction, which often induces a quantum phase transition and changes the sign of $R_\textrm{H}$ near the optimal doping for electron-doped cuprates \cite{Dagan2004evidence,li2019hole,barivsic2022high,Armitage2010progress}. As thus, a straightforward explanation of the $R_\textrm{H}$ discrepancy is the unique coexistence of the substituted Eu$^{3+}$ as electron dopants and the interstitial apical oxygens as hole dopants in the films, which can be best described by a two-band model with both electron-like and hole-like Fermi surfaces, as inserted in Fig.\ 2(b). In such a model \cite{Li2020superconducting, ziman1972principles}, the low-field $R_\textrm{H}$ denotes the difference between mobile electrons (by trivalent europium) and holes (by apical oxygens) as weighted by their mobilities, and still follows the $\mu_0$$H$ in a linear manner, as observed. Similar behaviors have been widely observed in various multiband superconductors, including the recently discovered infinite-layer nickelates sharing the same structure as SECO \cite{Li2020superconducting, Zeng2020phase, Osada2020Phase, zeng2022superconductivity}. More importantly, this explanation establishes a direct correlation between the ambipolar doping and $T_\textrm{c}$, i.e., the superconductivity benefits from a better match in density between the electrons and holes (small $R_\textrm{H}$ in magnitude) [Fig.\ 2]. With the introduction of apical oxygens in the EHC samples, the hole doping induces a downward shift in $E_\textrm{F}$, optimizing the matching between the two electron-like and hole-like pockets. The picture holds true as well for the ED samples as one compares the dependences of $T_\textrm{c}$ and $R_\textrm{H}$ on the Eu content $x$ in Figs.\ 1(e) and 2(b), aligning with our STM observation of an increase in apical oxygens as the Eu doping level rises \cite{Yu2022percolative}. Here, dominant electron doped by Eu$^{3+}$ positions the $E_\textrm{F}$ close to the upper Hubbard band, leading to relatively poor matching between the electron-like and hole-like pockets near $E_\textrm{F}$. These findings support that superconductivity benefits from the improved matching of electron and hole densities in both ED and EHC SECO samples.

\begin{figure}[h]
	\includegraphics[width=\columnwidth]{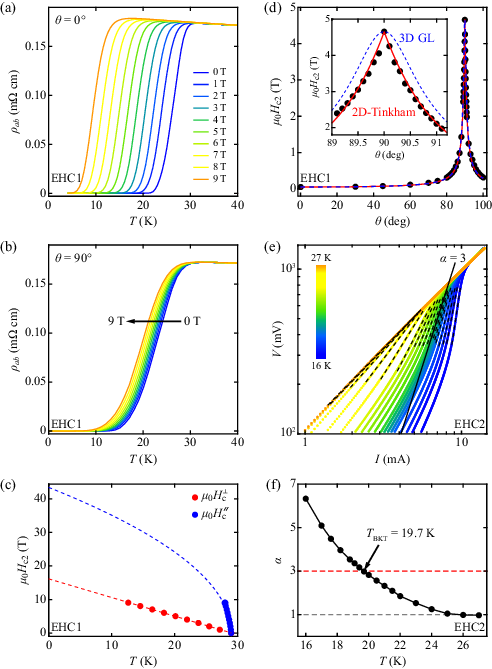}
	\caption{(a, b) In-plane resistivity $\rho_{ab}$ in the EHC1 ($x\sim$ 0.133) sample as a function of temperature $T$ under various magnetic fields perpendicular ($\theta$ = 0$^\textrm{o}$) and parallel ($\theta$ = 90$^\textrm{o}$) to the sample surface, respectively. (c) Temperature dependence of the upper critical field perpendicular ($\mu_0$$H_\textrm{c2}^\perp$) and parallel ($\mu_0$$H_\textrm{c2}^{\parallel}$) to the sample surface. The dashed lines show the best fits of the $\mu_0$$H_\textrm{c2}$($T$) to 2D GL equations. (d) Angular dependence of the $\mu_0$$H_\textrm{c2}(\theta)$ at 25 K. The inset shows a magnified region around $\theta$ = 90$^\textrm{o}$. The red and blue dashed lines are the theoretical fits of the $\mu_0$$H_\textrm{c2}(\theta)$ to 2D Tinkham formula $(H_\textrm{c2}(\theta)\textrm{sin}(\theta)/H_\textrm{c2}^{\parallel})^2$ + $|H_\textrm{c2}(\theta)\textrm{cos}(\theta)/H_\textrm{c2}^\perp|$ = 1 and the 3D anisotropic GL formula $(H_\textrm{c2}(\theta)\textrm{sin}(\theta)/H_\textrm{c2}^{\parallel})^2$ + $(H_\textrm{c2}(\theta)\textrm{cos}(\theta)/H_\textrm{c2}^\perp)^2$ = 1, respectively. (e) Voltage-current ($V$-$I$) curves of the EHC2 ($x\sim$ 0.141) sample plotted on a log-log scale. The black dashed lines show the fits to the power laws, $V$ $\varpropto$ $I^\alpha$, while the black solid line corresponds to the curve $V$ $\varpropto$ $I^3$ at the BKT transition temperature $T_{\textrm{BKT}}$. (f) Temperature dependence of the exponent $\alpha$ extracted from the power-law fits in (e). The intersection with the red dashes $\alpha$ = 3 gives a $T_{\textrm{BKT}} \sim$ 19.7 K.
	}
\end{figure}

Next, we turn to investigate angle $\theta$-resolved magnetoresistance in various SECO samples, with $\theta$ marking the angle between the external magnetic field $\mu_0$$H$ and the $c$ axis of the SECO films. Figures 3(a) and 3(b) exhibit temperature-dependent $\rho_{ab}$ in the EHC sample with $x\sim$ 0.133 (hereafter dubbed as EHC1), for $\mu_0$$H$ applied perpendicular ($\theta$ = 0$^\textrm{o}$) and parallel ($\theta$ = 90$^\textrm{o}$) to the sample surface, respectively. Apparently, the superconductivity is more dramatically suppressed by the out-of-plane magnetic fields than the in-plane ones. Such anisotropies are most probably indicative of a 2D nature of the superconductivity, in which the vortex motion is more remarkable in the out-of-plane magnetic field geometry. Indeed, Fig. 3(c) elaborates the temperature dependencies of the extracted upper critical field $\mu_0$$H_\textrm{c2}$ at $\theta$ = 0$^\textrm{o}$ ($\mu_0$$H_\textrm{c2}^{\parallel}$) and $\theta$ = 90$^\textrm{o}$ ($\mu_0$$H_\textrm{c2}^\perp$), which follows nicely the phenomenological Ginzburg-Landau (GL) formula for 2D superconducting films (see the dashed lines) \cite{tinkham2004introduction, kozuka2009two}, namely
\begin{equation}
\mu_0H_\textrm{c2}^\perp = \frac{\Phi_0}{2\pi\xi_{\textrm{GL}}^2(0)}(1-\frac{T}{T_\textrm{c}})
\end{equation}
\begin{equation}
\mu_0H_\textrm{c2}^\parallel = \frac{\sqrt{12}\Phi_0}{2\pi\xi_{\textrm{GL}}(0)d_{\textrm{SC}}}\sqrt{1-\frac{T}{T_\textrm{c}}}
\end{equation}
where $\Phi_0$ is a single magnetic flux quantum, $\xi_{\textrm{GL}}(0)$ is the extrapolated GL coherence length at $T$ = 0 K, and $d_{\textrm{SC}}$ is the temperature-independent superconducting thickness. We reveal that the 2D nature is generic to all ED and EHC SECO samples [Supplementary Fig.\ S3] \cite{supplementary}, giving rise to similar superconducting parameters from the GL fits above, as summarized Table 1. The superconductivity anisotropy parameter $\gamma$ = $H_\textrm{c2}^{\parallel}/H_\textrm{c2}^\perp$ is estimated as 3.4 $\pm$ 0.7 at 0 K, lying between those of nickelates and other more highly layered cuprates \cite{Mart1992manetic, sun2023evidence, Wei2023large}. Notably, the estimated superconducting thickness $d_\textrm{SC}$ $\sim$ 5.2 nm appears to be smaller than the film thickness $\sim$ 16.5 nm but much larger than $c$, which might correlate with the percolative nature of the 2D superconductivity in SECO films \cite{Yu2022percolative}.

In contrast to anisotropic 3D superconductors, the orbital pair-breaking effect in 2D superconductors is substantially diminished, thereby making the Pauli-limiting effect significantly more prominent. This fundamental distinction engenders a markedly different behavior of $H_\textrm{c2}(\theta)$. Figure 3(d) draws the angular dependence of $\mu_0$$H_\textrm{c2}(\theta)$, defined as the magnetic field where the resistivity $\rho_{ab}$ reaches 30\% of the normal-state $\rho_{ab}$ at 25 K, just below $T_\textrm{c}$ = 29.0 K. A cusp-like peak is consistently revealed near $\theta$ = 90$^\textrm{o}$ for all SECO samples [Supplementary Fig.\ S4] \cite{supplementary}. As expected, the $\mu_0$$H_\textrm{c2}(\theta)$ data follows the 2D Tinkham model better than the 3D anisotropic mass GL model \cite{tinkham2004introduction}, denoted by the red solid and blue dashed lines, respectively. This indicates that $H_\textrm{c2}$ is governed by the Pauli paramagnetic effect. Similar results have been recently found in the isostructural nickelates \cite{sun2023evidence, Wei2023large}, suggesting a universal 2D superconductivity for these infinite-layer oxide superconductors. This ends the long debate on the dimensionality of the superconductivity in these infinite-layer compounds \cite{sun2023evidence, Wei2023large,wang2021isotropic,kim2002three,Kim2002anisotropy,Zapf2005dimensionality}.

\begingroup
\squeezetable
\begin{table}[b]
	\caption{Comparison of the superconducting parameters, deduced from the angular magnetoresistance measurements, between the ED ($x\sim$ 0.141) and EHC SECO samples.}
	\centering
	\setlength\tabcolsep{1.2mm}
	\begin{footnotesize}
		
		\begin{tabular}{ c c c c c c }
			\hline
			& \multicolumn{1}{c}{$T_\textrm{c}$(K)} & \multicolumn{1}{c}{$\mu_0$$H_\textrm{c2}^\perp$(0)(T)} & \multicolumn{1}{c}{$\mu_0$$H_\textrm{c2}^{\parallel}$(0)(T)} & \multicolumn{1}{c}{$\xi_\textrm{GL}$(0)(nm)}& \multicolumn{1}{c}{$d_\textrm{SC}$(nm)} \\
			\hline \hline
			EHC1 & 29.0 & 16.0  & 43.1 & 4.9 & 5.8 \\
			\hline
			EHC2 & 25.6 & 13.1 & 43.4 & 5.0 & 5.2 \\
			\hline
			ED & 21.2 & 11.0  & 44.6 & 5.5 & 4.7 \\
			\hline \hline
		\end{tabular}
	\end{footnotesize}
\end{table}
\endgroup

\begin{figure}[t]
\includegraphics[width=\columnwidth]{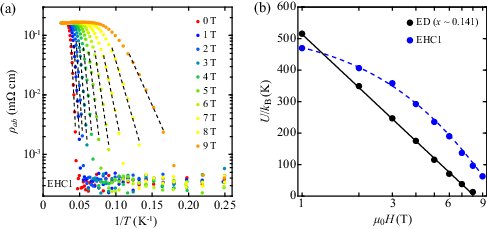}
\caption{(a) Arrhenius plots of $\rho_{ab}$ for various magnetic fields perpendicular to the sample surface. The black dashed lines denote the activated behavior of $\rho_{ab}$ close to $T_\textrm{c}$. (b) Extracted activation energy $U/k_\textrm{B}$ from the slopes of the dashed lines in (a), shown as a function of the magnetic field on a semilogarithmic plot. For comparison, the $U/k_\textrm{B}$ values in the ED sample with $x \sim$ 0.141 are shown in black circles, which follow the usual logarithmic evolution with the magnetic field (black line). The blue dashed line is a guide to the eye.
}
\end{figure}

We corroborate the 2D nature of the superconductivity by showing a clear-cut Berzzinsky-Kosterlitz-Thouless (BKT) transition in another EHC sample (EHC2, $x \sim$ 0.141), around which the zero-resistance state is driven by the binding of vortex-antivortex pairs. According to this model \cite{Beasley1979possibility,reyren2007superconducting}, the voltage scales with a power of current around $T_\textrm{c}$ in 2D superconductors ($V$ $\varpropto$ $I^\alpha$), with the exponent $\alpha$ = 3 at the BKT transition temperature $T_{\textrm{BKT}}$. As illustrated in Fig.\ 3(e), the SECO film indeed displays a well-defined BKT behavior and the exponent $\alpha$ amounts to three at $T_{\textrm{BKT}}$ = 19.7 K [Fig.\ 3(f)]. Near the BKT transition, the $\rho_{ab}$ turns out to be nicely fitted to the Halperin-Nelson equation [Supplementary Fig.\ S5] \cite{supplementary, reyren2007superconducting, Bardee1965theory}, which explains the energy dissipation from the Bardeen-Stephen vortex flow above $T_{\textrm{BKT}}$. This provides us an independent measure of the $T_{\textrm{BKT}}$ = 19.7 K that agrees excellently with the value deduced from the power-law analysis \cite{supplementary}.

Having established the 2D superconductivity and its invariance with the EHC, we explore the vortex dynamics in different SECO films. Shown in Fig.\ 4(a) are the Arrhenius plots of $\rho_{ab}$ for various $\mu_0$$H_\textrm{c2}^\perp$ in EHC1. Evidently, the temperature dependencies of $\rho_{ab}$ are characteristic of thermally activated behaviors just below $T_{\textrm{c}}$ described by $\rho_{ab}$ = $\rho_{0}$exp($-U(H)/k_\textrm{B}T$) (black dashed lines) with $k_\textrm{B}$ denoting the Boltzmann's constant. This is consistent with a thermally assisted vortex creep that leads to finite resistance in this regime \cite{Kundu2019effect}. Figure 4(b) compares two semilog plots of the extracted activation energy $U(H)$ $\sim H$ between the ED and EHC SECO samples, with more details shown in Supplementary Fig.\ S6 \cite{supplementary}. A logarithmic relationship is uniquely found between the $U(H)$ and $H$($U(H)\varpropto \textrm{ln}(H_0/H)$) for all ED samples, a hallmark of vortex creeping in two dimension \cite{Kundu2019effect,saito2015metallic,liao2018superconductivity}. However, the $U(H)$ in the EHC samples deviates hugely from such a logarithmic dependency on $\mu_0$$H$. It decreases more slowly with the magnetic field, which might explain the increased out-of-plane critical field $\mu_0$$H_\textrm{c2}^\perp$ in Table 1.

Our findings of the unique 2D superconductivity and its enhancement by the EHC in SECO films raise two significant questions on the infinite-layer cuprates. One is what kinds of Fermi pockets accommodate the coexisting electrons and holes for Cooper pairing. Recent scanning tunneling spectroscopic measurements revealed that the $E_\textrm{F}$ in the superconducting SECO films is well located within the charge-transfer gap (CTG) of the CuO$_2$ planes \cite{Yu2022percolative}. This implies that the Fermi pockets responsible for the superconductivity ought to be originated from the doping-induced low-energy bands within the CTG, which have been ever evidenced before \cite{zhong2020direct,fan2022direct,Harter2012Harter,Harter2015doping}. Remarkably, the low-energy bands were found to be composed of two distinct electron-like and hole-like pockets in surface electron-doped cuprates Ca$_3$Cu$_2$O$_4$Cl$_2$ \cite{hu2021momentum}, which are primarily derived from the upper Hubbard band and charge-transfer band, respectively. This matches with our experimental findings of the unique EHC in SECO. With increasing hole doping by the apical oxygens, the $E_\textrm{F}$ shifts downwards \cite{zhong2020direct,Molecular2020fan,Wang2020electronic,Yu2022percolative}, which can optimize the matching between the electron-like and hole-like Fermi pockets [Supplementary Fig.\ S7] \cite{supplementary} and therefore lowers $R_\textrm{H}$.

The other and more fundamental question regards to how the observed superconductivity and its enhancement link with the electron-like and hole-like pockets around $E_\textrm{F}$ in SECO. At first glance, our finding that the superconductivity benefits from the EHC implies that the inter-band interactions between the electron-like and hole-like pockets may be behind the pairing mechanism in SECO, just as proposed for the iron pnictides \cite{Unconventional2008Mazin,Unconventional2008Kuroki,hirschfeld2011gap,Kontani2010Orbital}. However, it should be noted that the emergent bands within the CTG are highly localized and strictly incoherent \cite{zhong2020direct,Wang2020electronic,Yu2022percolative,fan2022direct,Harter2012Harter,Harter2015doping,hu2021momentum,hu2021momentum}. The Fermi surface nesting scenario for itinerant carriers is most probably inapplicable to the present case. Although a further study is needed to fully understand this issue, we stress that the superconductivity and $T_\textrm{c}$ enhancement may benefit from a combined density of states from both electron-like and hole-like pockets in the framework of Bardeen-Cooper-Schrieffer theory \cite{suhl1959Bardeen}. This agrees with our recent observations of robust nodeless superconductivity and phonon excitations in infinite-layer cuprates \cite{fan2022direct,Yu2022percolative}.

In summary, we have demonstrated the 2D superconductivity in the SECO cuprate films. The revealed correlation between the superconductivity and EHC suggests the necessity of two-band models to understand the superconductivity in infinite-layer cuprates. This contradicts a common belief that single band models with only one type of carriers are sufficient to describe the electronic properties and superconductivity in cuprates. Our findings prominently resemble those of the infinite-layer nickelate superconductors and hold promise for a unified microscopic mechanism between the cuprate and multi-band iron-based superconductors.
\begin{acknowledgments}
The work was financially supported by the National Key R$\&$D Program of China (Grants No.\ 2022YFA1403100) and the Natural Science Foundation of China (Grants No.\ 12134008 and No.\ 52388201).

\end{acknowledgments}

\bibliography{SECO}

\end{document}